# Driver Age and Its Effect on Key Driving Metrics: Insights from Dynamic Vehicle Data

**Aparna Joshi**
*(Corresponding Author)*
Graduate Research Student
Department of Civil Construction and Environmental Engineering,
Iowa State University of Science and Technology
Ames, IA 50011-1066, USA
Email: aparnaj8@iastate.edu
Orcid Id: *0000-0002-7366-177X*

**Kojo Adugyamfi**
Graduate Research Student
Department of Civil Construction and Environmental Engineering,
Iowa State University of Science and Technology
Ames, IA 50011-1066, USA
Email: kgyamfi@iastate.edu

**Jennifer Merickel**
Health Science Administrator
Division of Behavioral and Social Research,
National Institute on Aging
Washington DC- Baltimore Area
E-mail: jennifer.merickel@nih.gov

**Pujitha Gunaratne**
Principal Scientist
Collaborative Safety Research Center,
Toyota Motor Engineering & Manufacturing North America Inc.,
Ann Arbor, MI, 48105
E-mail: pujitha.gunaratne@toyota.com

**Anuj Sharma**
Professor
Department of Civil Construction and Environmental Engineering,
Iowa State University of Science and Technology
Ames, IA 50011-1066, USA
E-mail: anujs@iastate.edu

## ABSTRACT

By 2030, the senior population aged 65 and older is expected to increase by over 50%, significantly raising the number of older drivers on the road. Drivers over 70 face higher crash death rates compared to those in their forties and fifties, underscoring the importance of developing more effective safety interventions for this demographic. Although the impact of aging on driving behavior has been studied, there is limited research on how these behaviors translate into real-world driving scenarios. This study addresses this need by leveraging Naturalistic Driving Data (NDD) to analyze driving performance measures - specifically, speed limit adherence on interstates and deceleration at stop intersections, both of which may be influenced by age-related declines. Using NDD, we developed Cumulative Distribution Functions (CDFs) to establish benchmarks for key driving behaviors among senior and young drivers. Our analysis, which included anomaly detection, benchmark comparisons, and accuracy evaluations, revealed significant differences in driving patterns primarily related to speed limit adherence at 75mph. While our approach shows promising potential for enhancing Advanced Driver Assistance Systems (ADAS) by providing tailored interventions based on age-specific adherence to speed limit driving patterns, we recognize the need for additional data to refine and validate metrics for other driving behaviors. By establishing precise benchmarks for various driving performance metrics, ADAS can effectively identify anomalies, such as abrupt deceleration, which may indicate impaired driving or other safety concerns. This study lays a strong foundation for future research aimed at improving safety interventions through detailed driving behavior analysis.

**INTRODUCTION**

Age-related driver performance changes are critical and require more research attention considering the growing prevalence of older drivers. By 2030, it is predicted that there will be a sharp rise in the number of senior citizens aged 65 and older, a rise of more than 50% (1). Senior drivers will become more prevalent overall on our roadways as a result. It is anticipated that by 2025, there will be 20% more licensed drivers in the United States who are 65 or older than there were in 2015 (2). This rise in the overall number of older drivers might impact traffic safety, particularly for those with medical issues. The abilities of some older drivers may be affected by age-related declines in vision, cognitive functioning (such as reasoning and memory), and physical changes. Numerous studies have indicated that the conditions such as early dementia, mild cognitive impairment, Parkinson's disease, and other neurodegenerative disorders, contribute to poorer driving performance, both in on-road evaluations (3, 4) and in simulated driving assessments (5), when compared to the overall older adult population. Additionally, compared to drivers in their forties and fifties (aged 35–54), drivers over the age of 70 have higher crash death rates per mile driven (6).

Past literature has studied age-related differences in driver behavior using various performance measures. Speed limit adherence is an essential performance metric which is used to track changes in speed in relation to each location's speed limit. Most studies used this essential performance metric to evaluate senior drivers' driving ability (7, 8). Several studies have compared the driving abilities of older and younger drivers based on age, and other research has examined the factors that contribute to senior drivers' driving errors (7,8). A study stated that even when older drivers exhibit better adherence to speed limits than younger drivers, the former group is at a higher risk of having a collision (9). Additionally, it is noticed that collisions are linearly proportional to reduction in speed compliance for all types of traffic densities (10). The age-gender interactions and the age group differences were found to exist for variables such as acceleration and speed related variables (11). The study categorized older drivers as "slow drivers" and younger drivers as "fast drivers". Along similar lines, a cross-sectional study conducted in California compared the differences in driving performance and reaction times for younger and older drivers (12). Their study concluded that older drivers have slower reactions, had more collisions, drove, deviated less in speed, and maintained a constant distance behind the car, in contrast to their younger counterparts. Another study observed that older drivers aged 65 and above showed poorer performance in several driving tasks, including reaction times, decision-making, and motor coordination, when compared to younger drivers. Interestingly, the younger groups, including teenagers and young adults, were found to commit more speeding errors than older drivers who were in good health (13). In addition, older participants reported observing significantly fewer road signs and exhibited a tendency to drive at slower speeds compared to younger participants (14). Furthermore, findings from a simulator study indicate that older drivers, regardless of their vision status, exhibited heightened eye movements and maintained slower driving speeds compared to younger drivers. (15).

Traffic safety and flow can be impacted by deceleration behavior, which is a crucial component of driving performance. Younger drivers, influenced by overconfidence and inexperience, often demonstrate riskier deceleration behaviors (16). A meta-analysis study highlighted the role of sensation-seeking and invulnerability in young drivers' risk-taking behaviors, influencing their deceleration decisions (17). Another study investigated the driving patterns across age and gender groups and found unexpected deceleration behavior, with older females showing lower rates over longer distances during deceleration. Controlling for start-stop speeds revealed older drivers had higher acceleration rates over shorter distances to reach equal speeds as younger drivers. Notably, older females displayed distinct deceleration patterns, exhibiting lower rates over extended distances compared to younger females (12).

Older drivers often struggle with lane changes, increasing their crash risk compared to other age groups (18). Analyzing 393 lane changes, it was found that older drivers frequently neglect over-the-shoulder glances before changing lanes, with failure rates of 98% for left and 92% for right lane changes (19). Additionally, older drivers show significantly slower response times in brake pedal tasks compared to younger drivers (20). They are also more likely to be involved in intersection crashes, particularly when making left turns (21). A study using data mining on 5-year crash data from Tokyo identified factors associated with crossing crashes involving older drivers, such as crash location, lighting, road and weather conditions, vehicle type, and traffic violations (22). In a naturalistic observation, older drivers (≥74 years) were compared to middle-aged drivers (35–64 years) based on their driving environment and error rates. Results showed that while both groups made few errors, older drivers preferred simpler routes and made fewer errors overall (23).

A systematic review investigated how cognitive decline affects driving simulator performance in older adults, noting increased variability and inconsistency in their driving (1). Another study involving 80 drivers aged 70 and older linked the Useful Field of View (UFOV) test to on-road driving performance, identifying common issues like "driving too fast," manual gearbox difficulties, and challenges with road signs and traffic lights (24). Research comparing distracted driving between middle-aged and older drivers found that the older drove slower, had less speed variability, and tended to "freeze up" under distraction (25). Additionally, driving performance deteriorates progressively from ages 20 to 80, with increased reaction times and variability, especially in complex tasks (26). A study on naturalistic driving found that older drivers identified fewer landmarks, drove slower, and made more safety errors compared to middle-aged drivers (27). It also revealed that older drivers focused more on road markings, while younger drivers paid more attention to dynamic objects like other cars (28). Studies of older drivers with rapid deceleration events showed slightly better vision and cognition measures (29), and the use of potentially inappropriate medications (PIMs) was linked to a higher risk of hard braking events, with increased risk associated with using multiple PIMs (30).

While past studies have found significant age-related differences in older driver performance, like speed limit adherence, acceleration/deceleration characteristics, and other driving measures, to date there has been little research to examine these differences in response to real-world road risks, especially using Naturalistic Driving Data (NDD). Such research is crucial for enhancing evidence-based insights aimed at informing older drivers about education programs, licensure criteria, and the design of safer transportation systems.

The aim of this study is to address the existing research gap by leveraging NDD to analyze and quantify key driving behaviors of both senior and younger drivers. By establishing baseline curves and percentile ranges for essential driving metrics—such as adherence to speed limits on interstates and deceleration at stop intersections—our approach aims to enhance the understanding of age-specific driving patterns. This paper proposes an approach that not only shows promising results but also informs and improves the development of Advanced Driver Assistance Systems (ADAS), ultimately contributing to more effective road safety interventions.

The paper is organized as follows: First, we outline the formulation of Key Performance Indices (KPIs) for evaluating driving behaviors. Next, we describe the dataset and sample, providing context for the analysis. We then explain the analytical methods and framework used to develop baseline curves and percentile ranges. Finally, we present key findings on driving behavior differences between senior and young drivers, followed by a discussion of implications, limitations, and future research avenues.

## FORMULATION OF KEY PERFORMANCE INDICES

The selection of Key Performance Indices involved a comprehensive approach aimed at evaluating driving performance for senior drivers. The process began with a detailed literature review and consultations with rehabilitation centers in Nebraska, specializing in senior driver assessment. To gather insights into standard measurements and assessment criteria, multiple Division of Motor Vehicles (DMV) driving performance score sheets were also reviewed. Subsequently, a questionnaire was developed based on the gathered information to identify critical factors for assessing senior drivers' driving capabilities. This questionnaire was circulated among driving rehabilitation occupational therapists, including discussions with experts from driving rehabilitation centers. Through these interactions, we gained valuable insights into their scoring methodologies and the effectiveness of different assessment factors.

Based on these inputs and discussions, the list of KPIs was refined to ensure it captured the essential metrics for evaluating driving performance among senior drivers. **Table 1** provides a detailed overview of the KPIs, incorporating insights from literature, DMV driving performance standards, and feedback from professional consultations. Important KPIs are also highlighted in the table, which were selected based on their commonality between DMV driving performance scoresheets and experts from driving rehabilitation centers. In our study, we focused on selecting KPIs based on their feasibility within our dataset. After careful consideration, we chose to prioritize speed limit adherence and deceleration as the primary metrics for evaluating driving performance among senior drivers. These KPIs were selected due to their availability and reliability within our dataset, aligning closely with DMV standards and expert recommendations.

**TABLE 1 Comprehensive List of KPIs from Experts at Rehabilitation Center and DMV Scoresheets**

| KPIs from Experts at Rehabilitation Center | KPIs from DMV Scoresheets | Important KPIs |
|---|---|---|
| | | |
| **Positioning** | | |
| Centers car in lane | - | ✓ |
| Safe following distance | Follows too closely | ✓ |
| Lane selection after turn | Gets into proper lane after turn | |
| Stopped in proper position | - | |
| | | |
| **Braking/Acceleration/Speed** | | |
| *Stopped smoothly at intersections\** | Smooth stop | ✓ |
| *Smooth acceleration from a stopped position \** | - | |
| Proper speed at intersections | Poor speed judgement approaching or at intersections | ✓ |
| *Adequate speed for conditions\** | Speed in ideal conditions (good weather, straight road, light traffic) | ✓ |
| Good regulation of gas or brake pedals | - | |
| | | |
| **Right and Left Turns** | | |
| Gradually applies brakes | Delayed or abrupt braking | |
| Uses turn signal | Makes proper signal of intention to turn | |
| Cancels signal | Cancel signal | |
| Correct lane selection after turn | Select proper lane after turn | ✓ |
| Yielded right of way | Yields without hesitating when necessary | |
| No lane changes middle of the intersection | | |
| Controlled steering and wheel recovery | Poor steering control | ✓ |
| | | |
| **Traffic Signal: Traffic Sign Compliance** | | |
| - | Deceleration rate approaching stop sign | |
| - | Heavy Deceleration rate approaching red light | ✓ |
| - | Heavy acceleration rate on green | ✓ |
| - | Red light running | ✓ |
| - | Waits too long after light turn green | |
| | | |
| **Stop Sign: Traffic Sign Compliance** | | |
| - | Proportion of stop sign violations | ✓ |
| - | Waits too long without any vehicle present | |
| - | Acceleration rate departing stop sign | ✓ |
| - | Look both way prior to stop sign | ✓ |

| | Small gap acceptance at stop sign (two-way) | |
| - | Small gap acceptance at permissive left (two-way) | |
| - | Waits too long at permissive left without any vehicle present | |
| | | |
| **Lane Change in Highway** | | |
| Checked mirror and blind spot | Lane change without blind spot check | ✓ |
| Showed good acceleration when lane changing | - | |
| Showed good steering control when lane changing | - | |
| Appropriately initiated lane changes | - | |
| Adjusted speed to flow of traffic | Too slow/impedes the traffic flow | |
| Signaled | - | |
| - | Number of lane violations per mile (Straight) | |
| - | Proportion of turning lane violation at signalized intersections | |
| | | |
| **Entering/Exiting in Highway** | | |
| Signaled entry/exit | - | |
| Checked blind spot | - | |
| Showed good acceleration | - | ✓ |
| Showed good steering control | - | |
| | | |
| **Traffic Conflict** | | |
| - | Tailgate - too close to the cars | ✓ |
| - | Swerving | ✓ |

**Note:** *denotes selected KPIs in our study*

## DATASET AND SAMPLE DESCRIPTION

### Baseline Development Dataset

NDD were extracted from existing registries at the University of Nebraska Medical Center (UNMC) containing 4–12 weeks of continuous, real-world driving from 2 age-cohorts: senior (N=35; 65–88 years of age) and young drivers (N=103; 21–64 years of age). Participants consented to study participation following UNMC IRB guidelines. All drivers were legally licensed and screened (self-report, clinical exam) to be typically aging with no major confounding medical conditions or medication usage. NDD were collected every second using sensors installed in each participant's own vehicle from on-off ignition (1 drive), collecting video (forward roadway, vehicle cabin) and vehicle sensor data (e.g., GPS, speed, accelerometer). GPS data were used to extract roadway speed limits from state GIS databases. Participants drove as typical. This dataset initially lacked posted speed limit information, necessitating the integration of speed data sourced from GIS databases and TIGER files.

This dataset was utilized to establish baseline Cumulative Distribution Function (CDF) curve for the selected KPIs, specifically focusing on speed limit adherence on interstates (65mph, 70mph, and 75mph) and stop intersection deceleration. The data underwent processing to meet the KPI requirements for creating CDF curves, which were used as reference or baseline distributions for further analyses. **Table 2** provides the summary and breakdown of the participants used for the baseline development for speed limit adherence. The dataset was specifically filtered to include only interstate roads and selected speed limits (≥65mph). Similarly, we also filtered the dataset to include participants who traveled through the stop intersections. **Table 3** provides a description of the participants included in the baseline development for deceleration at stop intersections.

**TABLE 2 Breakdown of the Processed Dataset for Speed Limit Adherence**

| Variables | Senior Category | Young Category |
|---|---|---|
| | | |
| Total Number of Datapoints | 12,151 | 23,418 |
| Total Number of Participants | 32 | 101 |
| | | |
| **Age of Participants** | | |
| Range | 65 - 70 | 21 - 64 |
| Mean | 67.8 | 49.3 |
| Standard Deviation | 1.08 | 12.7 |
| | | |
| **Sex of Participants** | | |
| Male | 19 | 53 |
| Female | 13 | 48 |
| | | |
| **No. of Participants in Various Speed Limits** | | |
| 75mph | 9 | 26 |
| 70mph | 24 | 78 |
| 65mph | 32 | 100 |

**TABLE 3 Breakdown of the Processed Dataset for Deceleration at Stop Intersection**

| Variables | Senior Category | Young Category |
|---|---|---|
| | | |
| Total Number of Datapoints | 14, 107 | 48, 544 |
| Total Number of Participants | 5 | 14 |
| | | |
| **Age of Participants** | | |
| Range | 65 - 70 | 39 - 64 |
| Mean | 66.8 | 54.3 |
| Standard Deviation | 1.85 | 6.79 |
| | | |
| **Sex of Participants** | | |
| Male | 2 | 8 |
| Female | 3 | 6 |

**Validation and Testing Dataset**

We employed a distinct naturalistic driving dataset consisting exclusively of senior drivers undergoing normal aging. This dataset was utilized to determine the optimal percentile range and subsequently test its effectiveness, as outlined in the results section. The study included a total of 47 participants aged between 66 and 88 years, with a mean age of 75 and a standard deviation of 5.47 years. Of the 47 participants, 15 were male and 32 were female. The existing dataset lacked posted speed limit information and hence, we prepared the data by finding the speed limit for all the drive points like baseline dataset. The data was cleaned for missing GPS, driving speed, and acceleration values. Further refinement was carried out by filtering the dataset to include only those traveling at 65mph and 75mph speed limits for adherence to speed limit KPI. Based on this filtered data, we selected validation and testing subsets to evaluate the performance of our model accurately.

**ANALYTICAL METHOD**

Kolmogorov-Smirnov (KS) tests were used to measure variability in the baseline CDF plots of senior and young drivers for both KPIs of interest. KS tests quantify the distance between either 1) the CDF of a reference and empirical distributions or 2) the empirical distribution of given two samples. It is a non-parametric test which compares the sample with the reference probability test (one-sample KS Test) or between two samples (two-sample KS test). The formula for the KS test can be given as **Equation 1**:

$$D = Max\ |F^{n1}(X) - F^b(X)|$$ (1)

where, n1 = Observations from a participant multiple drive
b = established baseline curve and,
D = KS statistic (maximum distance between two curves)

**METHODOLOGICAL FRAMEWORK**

The methodology employed in our study includes four major stages: 1) Data extraction and preparation, 2) Anomalies identification and removal, 3) Baseline curve development, and 4) Baseline curve validation.

**Speed Limit Adherence:** The data cleaning process involved removing missing data and filtering to focus on data points from interstate roads. This selection aimed to minimize confounding factors related to driver speed behaviors, such as traffic signals and stop signs. Anomalies were identified using CDFs across roadway segments, age groups, and participants within specified speed limits (75mph, 70mph, 65mph). Segments exhibiting anomalies were excluded, and baseline CDFs were established for both senior and young driver groups. These baseline CDF curves underwent validation to ensure accuracy and reliability.

**Stop Intersection Deceleration Behavior:** Deceleration behavior at stop intersections was analyzed using data from OpenStreetMap to identify and define intersections. Data was cleaned of points with empty coordinates and locations during preprocessing. A total of 75 intersections were used in this analysis. Each intersection was associated with buffer zones encompassing adjacent roadways where driver trajectories were tracked. Anomalies in these trajectories were identified and removed using CDFs, similar to the approach used for analyzing speed limit adherence.

*Data Extraction and Preparation*

For speed limit adherence, from the pool of data points, we segregated the data based on the age group i.e., senior participants (>=65) and young participants (65<). The whole dataset is filtered and cleaned for data points that lie only on interstate roads. To ensure a representative subset of each age cohort, we focused our study on interstate segments with sufficient driving points and participant data. Filtering out these sections was facilitated through a Tableau software dashboard (**Figure 1**). This led to a subsequent reduction in the dataset size, focusing on the most relevant and informative data points for the baseline development.

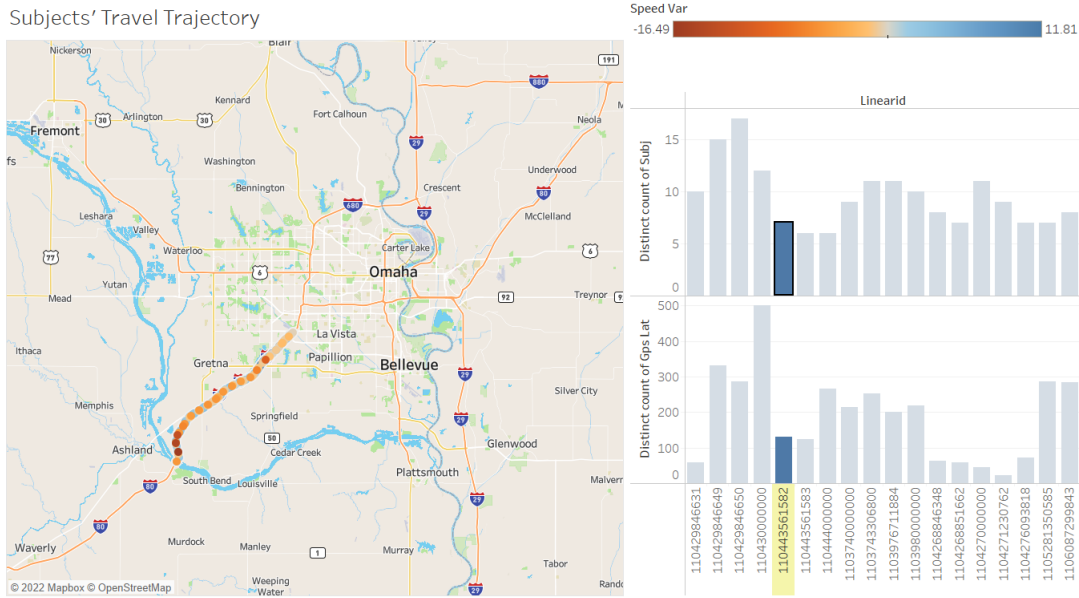

**Figure 1 Dashboard for selecting representative interstate segments based on driving points and participants**

Implementation of deceleration at the stop intersections in Omaha involved the use of stop locations from Open Street maps in Omaha. OpenStreetMap is a free, open-source mapping platform that provides detailed information on roads, intersections, and other geographic features. The approach involves two main steps: first, identifying stop locations using OpenStreetMap data and finding the roads that intersect at these intersections; secondly, creating buffers at these roads and identifying the trajectories which intersect with these buffers (**Figure 2**). Buffers are essentially areas around a particular location, in this case, the roads that intersect at the stop locations.

Vehicle deceleration was calculated at each time stamp of the driver's speed and time using the **Equation 2** below.

$$a = (v_2 - v_1) / t \qquad (2)$$

where:
- a is the deceleration (in meters per second squared, ft/s^2)
- $v_1$ is the initial speed (in meters per second, ft/s)
- $v_2$ is the final speed (in meters per second, ft/s)
- t is the time taken to decelerate (in seconds)

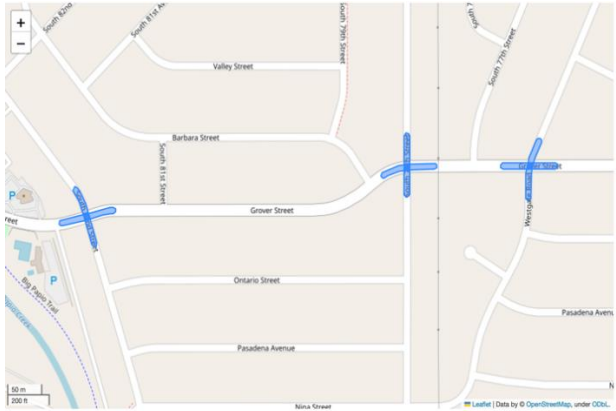

**(a) Buffered roads at stop intersections**

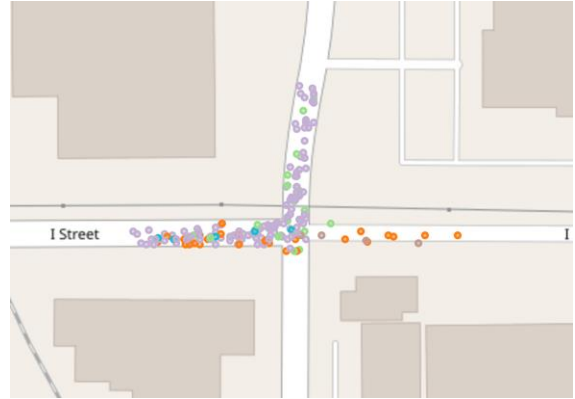

**(b) Filtered trajectories at stop intersection**

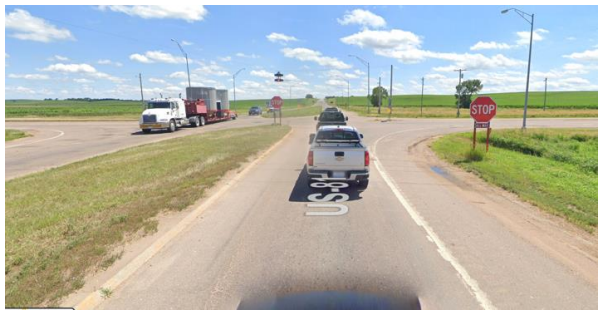

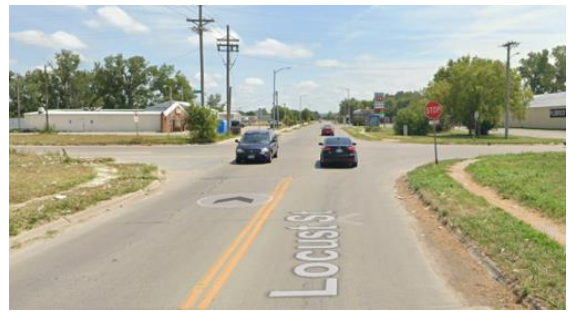

**(c) Typical extracted stop intersection**

**Figure 2 Approach to extract stop intersections**

*Anomalies Identification and Removal*

After identifying interstate road segments within specified speed limits (75mph, 70mph, and 65mph), CDFs were generated for both senior and young driver groups to evaluate adherence to regulatory speed limits. This section details the anomaly detection process specifically for senior drivers at 75mph. The same procedure was then applied to young drivers across varying posted speed limits.

**Figure 3 (a)** displays the plotted CDFs of interstate segments for senior drivers at 75mph. Two anomalies (non-overlapping CDFs) were observed: 110443561583 and 110444000000, identified based on road segments. An additional anomaly check was conducted at the participant level after removing these anomalous segments. Segments exhibiting speed distributions significantly different from the remaining segments were excluded from the analysis. Factors such as shoulder width and road geometry can influence speed behavior choices. Therefore, only segments with similar speed profiles were included to minimize the impact of unobserved variables on our model results. Consequently, segments 110443561583 and 110444000000 were excluded from further analysis.

Subsequently, we conducted an additional anomaly check focusing on participants. Participant SSS-DM-084 exhibited notable deviations in speed limit adherence compared to the other three participants, as evidenced in **Figure 3 (b).** Consequently, this participant was excluded from the baseline curve development process.

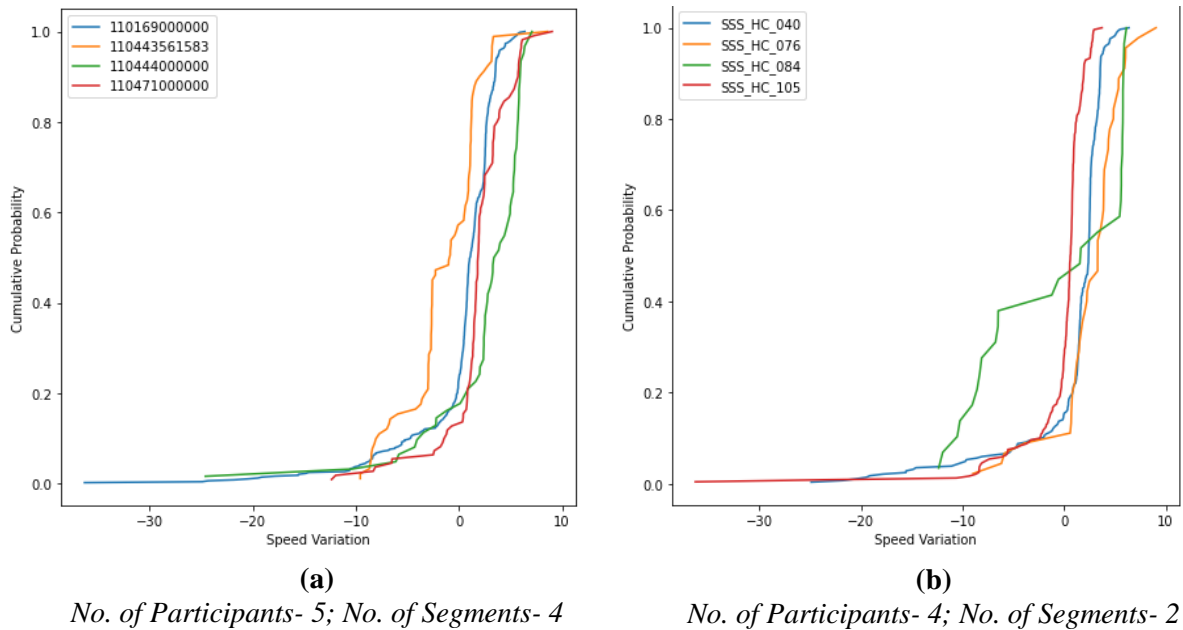

|  (a)  |  (b)  |
|---|---|
| *No. of Participants- 5; No. of Segments- 4* | *No. of Participants- 4; No. of Segments- 2* |

**Figure 3 CDF profile of adherence to 75mph speed limit by (a) road segment and (b) participant**

*Baseline Curve Development*

**Speed Limit Adherence:** In this step, baseline CDFs were developed by excluding participants and segments identified as anomalous in the previous step of anomaly identification. As a distinct baseline curve at 70mph among young drivers was not identifiable, our analysis proceeded to focus exclusively on the 75mph and 65mph speed limits. Baseline curves were derived for both age groups at 75mph and 65mph, as illustrated in **Figure 4** (a) and (b) respectively. For the 75mph speed limit, baseline curves were constructed using 2 segments and 3 participants for senior drivers, and 6 segments and 7 participants for young drivers. Similarly, for the 65mph speed limit, baseline CDFs were developed with 5 segments and 4 participants for senior drivers, and 4 segments and 15 participants for young drivers.

**Deceleration at Stop Intersection:** When comparing the baseline deceleration curves between senior and young drivers, **Figure 5** reveals a notable distinction. Young drivers exhibit CDF plots shifted to the left, indicating a higher occurrence of decelerations at higher values, whereas senior drivers show CDF plots shifted to the right, indicating a higher frequency of decelerations at lower values. This contrast can be attributed to several factors, including differences in driving experience, reaction times, and physical capabilities. Young drivers may tend towards risk-taking and aggressive driving behaviors, leading to more frequent and abrupt deceleration. In contrast, senior drivers, with potentially slower reaction times, often engage in more gradual and cautious deceleration.

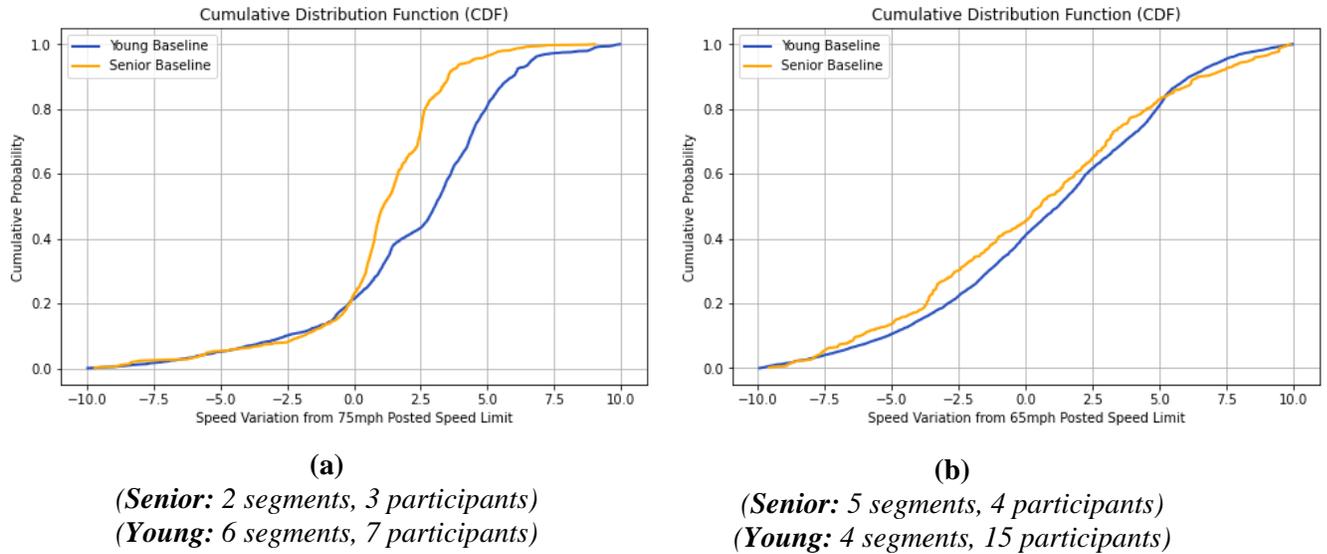

**(a)**

*(**Senior:** 2 segments, 3 participants)*
*(**Young:** 6 segments, 7 participants)*

**(b)**

*(**Senior:** 5 segments, 4 participants)*
*(**Young:** 4 segments, 15 participants)*

**Figure 4 Baseline curves for senior vs young drivers for (a) 75mph and (b) 65mph**

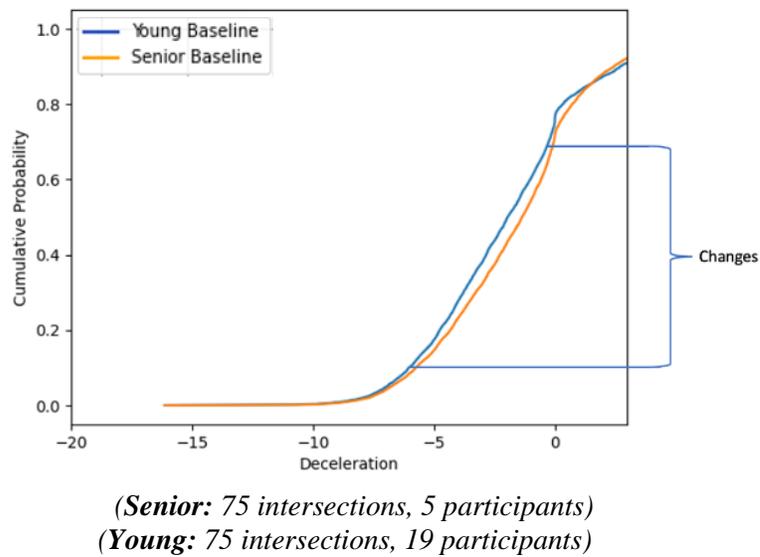

*(**Senior:** 75 intersections, 5 participants)*
*(**Young:** 75 intersections, 19 participants)*

**Figure 5 Baseline curves for senior vs young drivers for deceleration at stop intersection**

*Baseline Curve Validation*

We utilized the non-parametric Kolmogorov-Smirnov (KS) test to assess differences in speed limit adherence between senior and young drivers based on baseline curves at 75mph and 65mph. To refine and validate the speed metric, we identified 5 senior participants for each speed limit category (75mph and 65mph) from the testing dataset. This selection allowed us to pinpoint the percentile range that maximally differentiated each participant from the young baseline CDF and minimized the average distance from the senior baseline CDF. Detailed findings from this analysis are discussed in the subsequent section.

## RESULTS

### KS Test Results

**Table 4** shows the results of the KS test for the two KPIs. The findings indicate significant differences in speed limit adherence behavior between senior and young drivers at 75mph. However, contrasting outcomes emerge at the 65mph speed limit, suggesting no substantial disparities in speed limit adherence between these age groups. Typically, road segments with a 65mph speed limit are in more urban areas, which may have increased enforcement, leading to a similar risk-taking behavior between senior and young drivers. Conversely, 75mph segments are generally on freeways connecting cities. In these segments, younger drivers often exceed the speed limit more frequently, while older drivers display a comparable risk-taking attitude to that observed in urban freeway segments. Additionally, the results underscore distinct differences in deceleration behavior at stop intersections between senior and young drivers.

**TABLE 4 KS test results for selected KPIs**

| Senior vs Young Baseline | Test Statistic | Results |
|---|---|---|
| | | |
| For 75mph speed limit | KS-stat | **0.22844*** |
| | p-value | *< 2.2e-16* |
| For 65mph speed limit | KS-stat | 0.088853 |
| | p-value | *0.2264* |
| For deceleration at stop intersection | KS-stat | **0.06553*** |
| | p-value | *1.36e-10* |

**Note:** *Statistically significant at 95th percentile*

### Accuracy of Performance Metrics

*Optimization of Percentile Range Selection*

During this stage, our objective was to identify the optimal percentile range where the validation dataset closely aligns with the senior baseline curve. We aimed to extract senior participants from the validation and testing dataset who traveled on road segments with either a 75mph or 65mph speed limit. This approach resulted in 5 subjects for each speed limit category, ensuring that our refined classification metric was based on participants from the same road segments. Similarly, from the same dataset, we identified 5 subjects for the deceleration at the stop intersection metric who traveled through the same intersection. Next, the algorithm determined the percentile range that minimizes the distance to the senior baseline CDF while maximizing the distance to the young baseline CDF across all participants. Subsequently, we generated a scatterplot to visually depict the distances from the baselines within the selected percentile range. Participants in the validation dataset exhibiting abnormal deviations from the baseline curves were excluded based on visual inspection of CDFs. This precautionary step was crucial to mitigate potential biases during the percentile range selection process.

**75mph Speed Limit: Figure 6 (a)** illustrates the CDFs for 75mph created using validation data which consist of senior drivers on road segments distinct from those used to construct the baseline curves. This approach highlights variations observed between the baseline curves and the participants in the validation set. However, overarching characteristics, such as the tendency of a significant portion of senior drivers to adhere to speed limits, hold true for these selected participants (grey line CDFs in **Figure 6 (a)**). Among these 5 participants at 75mph, the identified percentile range was determined to be from the 68th to the 95th percentile. **Figure 6 (b)** displays the scatterplot representing the validation dataset in this percentile range. As we can see in **Figure 6 (b)**, points that fall below the red line indicate that these participants are closer to the senior baseline compared to the young baseline. To simplify, each point is represented by coordinates *(x, y)*, where *x* is the distance from the young baseline and *y* is the distance from the senior baseline. If a point lies below the line $y = x$, it means that $x$ is greater than $y$. In other words, the distance from the young baseline is greater than the distance from the senior baseline, suggesting that the participant's driving behavior aligns more closely with the senior driver category.

**65mph Speed Limit:** Similarly, **Figure 6 (b)** shows the CDFs developed using the validation dataset for 65mph speed limit. For these 5 participants within 65mph, the percentile range came out to be 36th to 58th percentile and the scatterplot of the validation dataset is shown in **Figure 6 (d)**.

**Deceleration at Stop Intersection: Figure 7 (a)** shows the CDFs developed using the validation dataset for deceleration at stop intersection. The results in **Figure 7 (b)** shows that for 5 participants, the percentile range spanning from the 35th to 70th percentile is the optimal bracket which notably maximizes the distance from the young baseline while simultaneously minimizes the distance to the senior baseline.

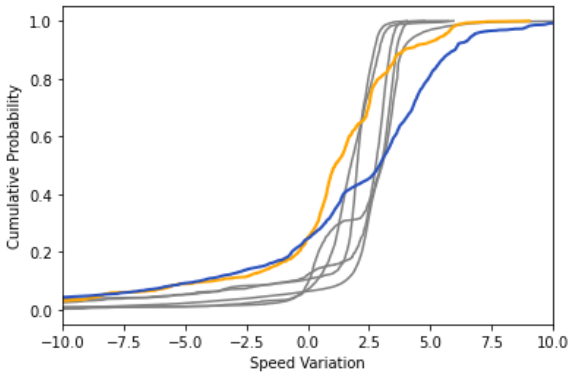

**(a) For 75mph**

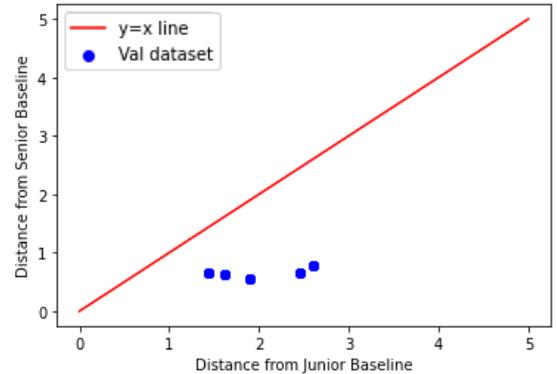

**(b) For 75mph**

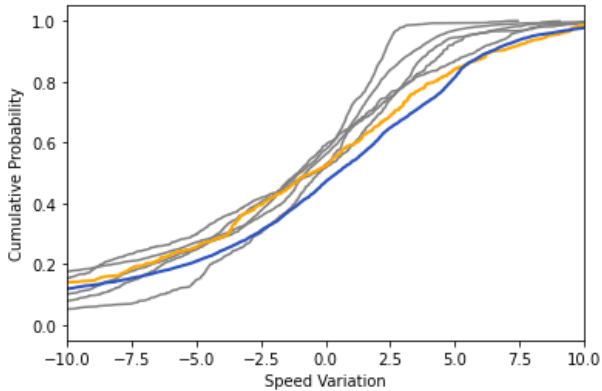

**(c) For 65mph**

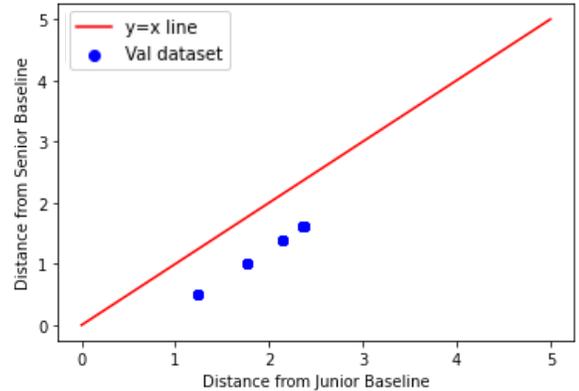

**(d) For 65mph**

**Figure 6 (a) and (c) CDFs of baseline curves and validation dataset; (b) and (d) Scatterplot of distances from baseline for identified percentile range for speed limit adherence.**

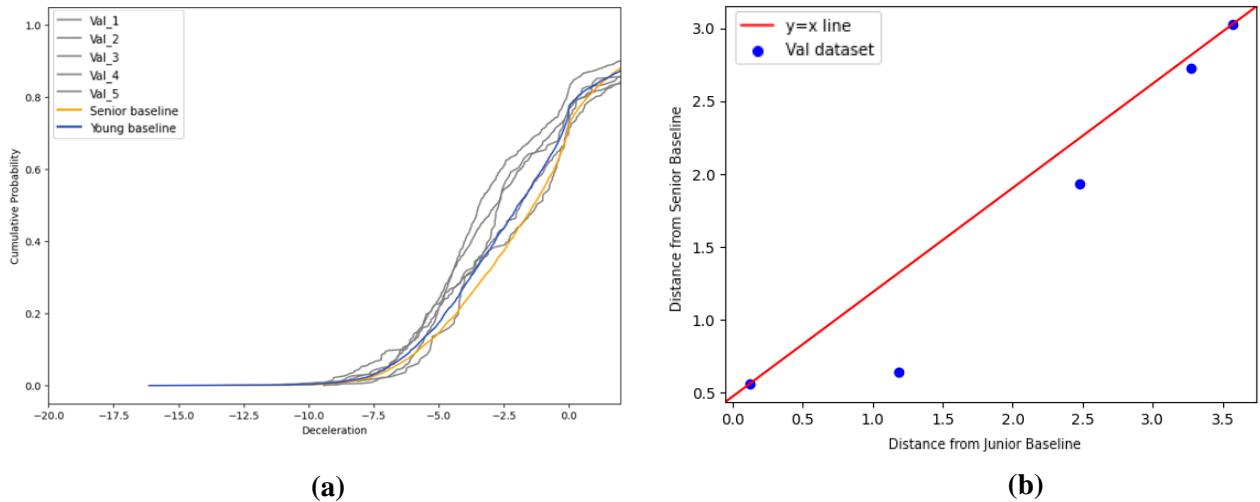

**(a)**                                                                                                     **(b)**

**Figure 7 (a) CDFs of baseline curves and validation dataset; (b) Scatterplot of distances from baseline for identified percentile range for deceleration at stop intersections**

*Checking the metric accuracy with Testing Dataset*

To evaluate the generalization of our metrics to new, unseen data, we filtered the validation and testing dataset to include only senior participants who traveled on the same road segments with 65mph or 75mph speed limits. This filtering process resulted in 18 participants being selected for evaluating speed limit adherence at 65mph and 75mph. After filtering for participants traveling through the same stop intersections, we identified 18 participants for the deceleration study. These participants were tested using the identified percentile range to determine their categorization as senior drivers.

**75mph Speed Limit:** As depicted in **Figure 8 (a),** data points below the red line indicate that these participants are closer to the senior baseline than the young baseline, while points above the line indicate the opposite. The results reveal that 14 out of 18 senior participants fall within the senior driver category, as they lie below the red line within the 68th to 95th percentile range. This classification achieves an accuracy of 77.8%. This finding, supported by statistically significant KS test results at the 95th percentile for the 75mph speed limit, suggests that adherence to higher speed limits may indicate age-related differences in driving behavior. These differences could be due to variations in cognitive processing speed, risk tolerance, and confidence levels among age groups.

**65mph Speed Limit: Figure 8 (b)** shows that 8 out of 18 senior participants fall within the senior driver category, as indicated by lying below the red line in the 36th to 52nd percentile range. This classification yields an accuracy rate of 44.4%. An accuracy rate of 44.4% indicates poor model performance, suggesting it's not much better than random guessing. This aligns with the KS test results, which found no statistical significance between senior and young baseline curves at 65mph.

**Deceleration at Stop Intersection: Figure 9** shows that 5 out of 12 senior participants aligned with the senior baseline curve in the 35th to 70th percentile range, achieving a 41.6% accuracy. Despite the KS test showing statistical significance between young and senior drivers' CDFs, this low accuracy suggests that differences in driving behaviors are not uniform across all percentiles. The overlap in driving behaviors within specific percentile ranges complicates precise classification. Additionally, lower accuracies may stem from a narrower distribution of behaviors in this metric, influenced by external factors like traffic conditions or road design, resulting in less distinct patterns between the age groups.

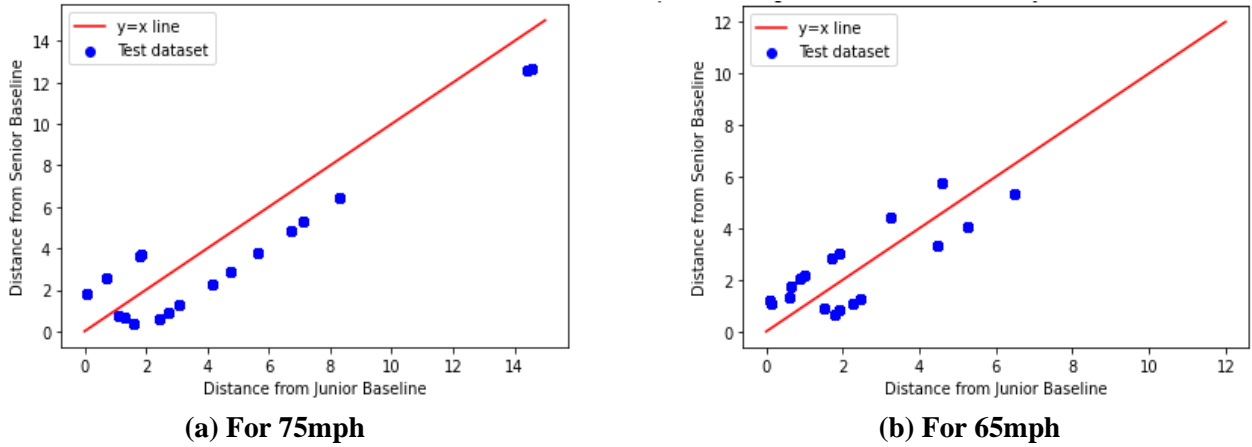

(a) For 75mph                     (b) For 65mph

**Figure 8 Scatterplot for testing dataset for speed limit adherence**

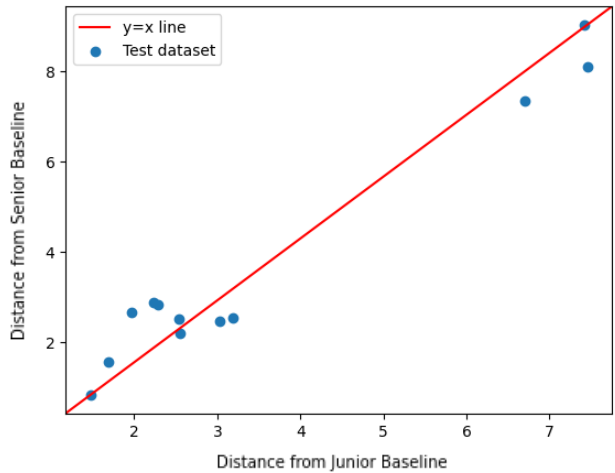

**Figure 9 Scatterplot for testing dataset for deceleration at stop intersection**

**CONCLUSIONS AND FUTURE AVENUES**

In this study, we proposed an approach that explored multiple KPIs to differentiate between young and older drivers and quantify their driving behavior. The selection of these KPIs was informed by a comprehensive approach, including a literature review, consultations with senior driver rehabilitation experts, and review of DMV driving performance score sheets. These sources helped to identify critical assessment factors, leading us to prioritize speed limit adherence and deceleration at stop intersections as the primary measures due to their relevance and reliability within our dataset.

To achieve this, we employed a methodology that involved extracting relevant naturalistic driving data, detecting anomalies, and developing CDFs for baseline comparisons of measures including speed limits at 75mph and 65mph, as well as deceleration at stop intersections. We evaluated these benchmarks to establish performance metrics and determine their effectiveness in distinguishing between senior and young drivers.

Our proposed approach identified adherence to the 75mph speed limit as the most effective measure for differentiating age-related driving behavior, with 77% accuracy based on the average distance from baseline curves between the 60th and 95th percentiles. This high accuracy suggests significant differences in cognitive processing, risk assessment, and confidence levels between age groups at higher speeds. Younger drivers may be more likely to exceed speed limits, indicating a higher risk tolerance, while older drivers may drive more cautiously due to age-related cognitive declines and slower reaction times. In contrast, for deceleration, the average distance from baseline curves between the 35th and 70th percentiles achieved only 41.6 % accuracy in distinguishing senior drivers. This lower accuracy could be attributed to the less variable nature of deceleration behavior, which may be more influenced by external factors such as traffic conditions, road design, or the necessity of coming to a stop, resulting in a narrower distribution of behavior across age groups. Similarly, adherence to the 65mph speed limit achieved only 44.4% accuracy within the 36th to 52nd percentile range, possibly due to more uniform compliance and reduced variability among drivers. While adherence to the 75mph speed limit is a promising indicator of age-specific driving characteristics, further data collection and analysis are needed to refine and validate metrics for other driving behaviors.

These findings have significant implications for enhancing Advanced Driver Assistance Systems and improving road safety across different age groups. By establishing precise baseline curves and identifying percentile ranges for key driving behaviors, our research enables the development of more personalized and effective ADAS features. The ability to differentiate between senior and young drivers based on metrics such as speed limit adherence and deceleration patterns allows for the implementation of targeted interventions. For instance, ADAS can offer adaptive feedback and alerts that are tailored to the specific driving characteristics of each age group. This might include more frequent speed limit reminders or smoother deceleration assistance for senior drivers, while younger drivers could receive different forms of guidance or warnings. Moreover, the refined benchmarks provided by our study support advanced anomaly detection capabilities. Deviations from established baselines can signal potentially hazardous driving behaviors, such as excessive speeding or erratic deceleration, which could indicate impaired driving or other safety concerns. By identifying these anomalies in real-time, ADAS can prompt timely corrective actions to prevent accidents and improve overall driving safety. Additionally, the continuous refinement of these baselines with new data will facilitate ongoing advancements in ADAS technologies, ensuring they remain effective as driving behaviors evolve. This approach not only enhances individual driver safety but also contributes to broader traffic safety improvements and regulatory compliance by providing a standardized method for assessing and addressing age-specific driving risks.

This paper acknowledges limitations that we urge researchers to address in subsequent studies. It is pivotal to note that the study's baseline development for speed limit adherence was limited by a small sample size, as the data was filtered to include only specific speed limit values. Consequently, a more comprehensive data set would be instrumental in drawing a more definitive conclusion on this matter. Given our current focus on uninterrupted traffic flow (i.e., interstates), there is an avenue for future research to study less uniform traffic environments like residential and commercial settings. This exploration would provide a more comprehensive understanding of the road classification dynamics. Subsequent investigations might delve into alternative driving key performance parameters, such as lane deviation, reaction time, and lateral/longitudinal deceleration patterns during interactions with cross traffic, to establish connections between age and driver safety. Furthermore, future research may study other potential factors that could influence deceleration rates, including variables like weather conditions, prevailing traffic density, and the specific nature of the intersection.

## ACKNOWLEDGMENTS

The authors express their gratitude to the Toyota Collaborative Safety Research Center for providing financial support for this research endeavor. Additionally, authors extend their appreciation to the University of Nebraska Medical Center for their assistance in the recruitment and management of research participants.

## AUTHOR CONTRIBUTIONS

The authors confirm contribution to the paper as follows: A. Joshi conceived the study, conducted the analysis, and drafted and finalized the manuscript. K. Adugyamfi contributed to the analysis. J. Merickle, facilitated dataset acquisition used in the study. P. Gururatne, from the funding agency, provided valuable insights into the analysis process and results. A. Sharma, as advisor, contributed to the study conception and provided guidance in fine-tuning the research throughout its development stages. All authors reviewed the results and approved the final version of the manuscript.

**REFERENCES**

1) Depestele, S., Ross, V., Verstraelen, S., Brijs, K., Brijs, T., Dun, K.V., and Meesen, R.L. The Impact of Cognitive Functioning on Driving Performance of Older Persons in Comparison to Younger Age Groups: A Systematic Review. Transportation Research Part F: Traffic Psychology and Behaviour, 2020, 73: 433-452.

2) National Highway Traffic Safety Administration. Traffic Safety Facts – Older Population 2015 Data. U.S. Department of Transportation, Washington, D.C., 2017.

3) Devos, H., Vandenberghe, W., Nieuwboer, A., Tant, M., De Weerdt, W., Dawson, J.D., and Uc, E.Y. Validation of a Screening Battery to Predict Driving Fitness in People with Parkinson's Disease. Movement Disorders, 2013, 28(5): 671-674.

4) Uc, E.Y., Rizzo, M., Anderson, S.W., Sparks, J., Rodnitzky, R.L., and Dawson, J.D. Impaired Visual Search in Drivers with Parkinson's Disease. Annals of Neurology, 2006, 60(4): 407-413.

5) Rizzo, M., McGehee, D.V., Dawson, J.D., and Anderson, S.N. Simulated Car Crashes at Intersections in Drivers with Alzheimer Disease. Alzheimer Disease & Associated Disorders, 2001, 15(1): 10-20.

6) Cox, A.E., and Cicchino, J.B. Continued Trends in Older Driver Crash Involvement Rates in the United States: Data Through 2017–2018. Journal of Safety Research, 2021, 77: 288-295.

7) Lee, H.C., Cameron, D., and Lee, A.H. Assessing the Driving Performance of Older Adult Drivers: On-Road Versus Simulated Driving. Accident Analysis & Prevention, 2003, 35(5), 797-803.

8) Dawson, J.D., Uc, E.Y., Anderson, S.W., Johnson, A.M., and Rizzo, M. Neuropsychological Predictors of Driving Errors in Older Adults. Journal of the American Geriatrics Society, 2010, 58(6): 1090-1096.

9) Mouloua, M., Rinalducci, E., Smither, J., and Brill, J. Effect of Aging on Driving Performance. Proceedings of the Human Factors and Ergonomics Society Annual Meeting, 2004, 48: 253-257.

10) Olmez, S., Douglas-Mann, L., Manley, E., Suchak, K., Heppenstall, A., Birks, D., and Whipp, A. Exploring the Impact of Driver Adherence to Speed Limits and the Interdependence of Roadside Collisions in an Urban Environment: An Agent-Based Modelling Approach. Applied Sciences (Switzerland), 2021, 11(12).

11) Nakagawa, S., Kriellaars, D., Blais, C., Montufar, J., and Porter, M.M. Speed and Acceleration Patterns of Younger and Older Drivers. Canadian Multidisciplinary Road Safety Conference, 2006.

12) Doroudgar, S., Chuang, H.M., Perry, P.J., Thomas, K., Bohnert, K., and Canedo, J. Driving Performance Comparing Older Versus Younger Drivers. Traffic Injury Prevention, 2017, 18(1): 41-46.

13) Carr, D., Jackson, T.W., Madden, D.J., and Cohen, H.J. The Effect of Age on Driving Skills. Journal of the American Geriatrics Society, 1992, 40(6): 567-573.

14) Chaparro, A., Wood, J.M., and Carberry, T. Effects of Age and Auditory and Visual Dual Tasks on Closed-Road Driving Performance. Optometry and Vision Science, 2005, 82(8): 747-754.

15) Szlyk, J.P., Seiple, W., and Viana, M. Relative Effects of Age and Compromised Vision on Driving Performance. Human factors, 1995, 28(1): 430-436.

16) Deery, H.A. Hazard and Risk Perception Among Young Novice Drivers. Journal of Safety Research, 1999, 30(4): 225-236.

17) Özkan, T., and Lajunen, T. Multidimensional Traffic Locus of Control Scale (T-LOC): Factor Structure and Relationship to Risky Driving. Personality and Individual Differences, 2005, 38(3): 533-545.

18) Chandraratna, S., and Stamatiadis, N. Problem Driving Maneuvers of Elderly Drivers. Transportation Research Record, 2003, 1843: 89-95.

19) Antin, J.F., Wotring, B., Perez, M.A., and Glaser, D. Investigating Lane Change Behaviors and Difficulties for Senior Drivers Using Naturalistic Driving Data. Journal of Safety Research, 2020, 74: 81-87. DOI: 10.1016/j.jsr.2020.04.008.

20) Shin, H., and Lee, H. Characteristics of Driving Reaction Time of Elderly Drivers in the Brake Pedal Task. Journal of Physical Therapy Science, 2012, 24: 567-570.

21) Mayhew, D.R., Simpson, H.M., and Ferguson, S.A. Collisions Involving Senior Drivers: High-Risk Conditions and Locations. Traffic Injury Prevention, 2006, 7(2): 117-124. DOI: 10.1080/15389580600636724.

22) Yang, J., Higuchi, K., Ando, R., and Nishihori, Y. Examining the Environmental, Vehicle, and Driver Factors Associated with Crossing Crashes of Elderly Drivers Using Association Rules Mining. Journal of Advanced Transportation, 2020, 2019. DOI: 10.1155/2020/2593410.

23) Mazer, B., Chen, Y.T., Vrkljan, B., Marshall, S.C., Charlton, J.L., Koppel, S., and Gélinas, I. Comparison of Older and Middle-Aged Drivers' Driving Performance in a Naturalistic Setting. Accident Analysis & Prevention, 2021, 161: 106343. DOI: 10.1016/j.aap.2021.106343.

24) Dukic Willstrand, T., Broberg, T., and Selander, H. Driving Characteristics of Older Drivers and Their Relationship to the Useful Field of View Test. Gerontology, 2017, 63(2): 180-188. DOI: 10.1159/000448281.

25) Thompson, K.R., Johnson, A.M., Emerson, J.L., Dawson, J.D., Boer, E.R., and Rizzo, M. Distracted Driving in Elderly and Middle-Aged Drivers. Accident Analysis & Prevention, 2012, 45(2): 711-717. DOI: 10.1016/j.aap.2011.09.040.

26) Svetina, M. The Reaction Times of Drivers Aged 20 to 80 During a Divided Attention Driving. Traffic Injury Prevention, 2016, 17(8): 810-814. DOI: 10.1080/15389588.2016.1157590.

27) Aksan, N., Dawson, J.D., Emerson, J.L., Yu, L., Uc, E.Y., Anderson, S.W., and Rizzo, M. Naturalistic Distraction and Driving Safety in Older Drivers. Human Factors, 2013, 55(4): 841-853. DOI: 10.1177/0018720812465769.

28) Dukic, T., Broberg, T., and Broberg, T. Older Drivers' Visual Search Behaviour at Intersections. Transportation Research Part F: Traffic Psychology and Behaviour, 2012, 15: 462-470.

29) Keay, L.J., Munoz, B., Duncan, D.D., Hahn, D.V., Baldwin, K.C., Turano, K.A., Munro, C.A., Bandeen-Roche, K., and West, S.K. Older Drivers and Rapid Deceleration Events: Salisbury Eye Evaluation Driving Study. Accident Analysis & Prevention, 2013, 58: 279-285.

30) Xue, Y., Chihuri, S., Andrews, H.F., Betz, M.E., DiGuiseppi, C., Eby, D.W., Hill, L.L., Jones, V., Mielenz, T.J., Molnar, L.J., Strogatz, D., Lang, B.H., Kelley-Baker, T., and Li, G. Potentially Inappropriate Medication Use and Hard Braking Events in Older Drivers. Geriatrics, 2021, 6(1): 20. DOI: 10.3390/geriatrics6010020.